\documentclass[fleqn,usenatbib,twocolumn]{mnras}
\usepackage[T1]{fontenc}
\usepackage{ae,aecompl}
\usepackage[dvips]{graphicx}	
\usepackage{amsmath}	
\usepackage{amssymb}	
\usepackage{longtable,booktabs,caption,threeparttablex}
\newcommand{\kms}{\,km\,s$^{-1}$} 

\title{A study of Si II and S II features in spectra of Type Ia supernova}

\author[X. Zhao et al.]{
Xulin Zhao$^{1}$\thanks{E-mail: zhaoxulin@139.com (TJUT)},
Keiichi Maeda$^{2}$,
Xiaofeng Wang$^{3}$
and Hanna Sai$^{3}$
\\
$^{1}$School of Science, Tianjin University of
Technology, Tianjin, 300384, China\\
$^{2}$Department of Astronomy, Kyoto University,
Kitashirakawa-Oiwake-cho, Sakyo-ku, Kyoto 606-8502, Japan\\
$^{3}$Physics Department and Tsinghua Center for
Astrophysics, Tsinghua University, Beijing, 100084, China\\
}

\date{Accepted XXX. Received YYY; in original form ZZZ}

\pubyear{2020}

\begin{document}
\label{firstpage}
\pagerange{\pageref{firstpage}--\pageref{lastpage}}
\maketitle

\begin{abstract}
We studied the spectral features of Si II $\lambda\lambda$4130, 5972, 6355 and S II W-trough for a large sample of Type Ia supernovae (SNe Ia). We find that in NV (Normal-Velocity) subclass of SNe Ia, these features tend to reach a maximum line strength near maximum light, except for Si II $\lambda$5972. Spectral features with higher excitation energy, such as S II W-trough, are relatively weak and have relatively low velocity. SNe Ia with larger $\Delta$m$_{15}$($B$) tend to have lower velocities especially at phases after maximum light. NV SNe show a trend of increasing line strength with increasing $\Delta$m$_{15}$($B$), while 91T/99aa-like SNe show an opposite trend. Near maximum light, the absorption depth of Si II $\lambda$5972 shows the strongest correlation with $\Delta$m$_{15}$($B$), while at early times the sum of the depths of Si II $\lambda\lambda$4130 and 5972 shows the strongest correlation with $\Delta$m$_{15}$($B$). The overall correlation between velocity and line strength is positive, but within NV SNe the correlation is negative or unrelated. In normal SNe Ia, the velocity-difference and depth-ratio of a longer-wavelength feature to a shorter-wavelength feature tend to increase with increasing $\Delta$m$_{15}$($B$). These results are mostly well explained with atomic physics, but some puzzles remain, possibly related to the effects of the saturation, line competition or other factors.

\end{abstract}


\section{Introduction\label{S1}}

Type Ia supernovae (SNe Ia) remain the best distance indicators on cosmological scales \citep{Phillips93, Phillips99, Riess98, Riess16, Riess19, Perlmutter99, Betoule14, Dhawan18, Jones18, Scolnic18, Freedman19}. However, their progenitor systems \citep[e.g.,][]{Whelan73, Nomoto82, Nomoto97,Iben84, Webbink84, Maoz14, Cao15, Olling15, Maeda16, Blondin17, Liu18, Livio18, Jha19, Flors20} and the explosion mechanisms \citep{Nomoto84, Nomoto18, Khokhlov91, Hillebrandt00,Maeda10,Maeda18, Sim12, Sim13, Ruiter13, Shen14, Jha19, Wu20} remain unclear. Spectroscopic information is a key to unraveling the persistent mysteries. Many physical properties, such as the element distributions, the kinematics of the ejecta, the photospheric temperature, and the optical depth can only be determined by spectroscopic methods \citep[e.g.,][]{Sternberg11, Dilday12, Parrent14, Maguire18, Wilk18, Jacobson19}.


SN Ia spectra are characterized by strong absorption features of intermediate mass elements (IME) ions, such as Si II, S II and Ca II \citep[see][for a comprehensive review]{Filippenko97}. Of particular interest is the prominent feature Si II $\lambda$6355. The blueshifted velocity of the photospheric-velocity feature (PVF) of Si II $\lambda$6355 is representative of the ejecta velocity at the photosphere, typically about 11,000 km s$^{-1}$ at $B-$band maximum \citep{Benetti05,Maeda10,Wang09,Wang13}. In some early spectra, there is a high velocity feature (HVF) of Si II $\lambda$6355, detached from the PVF, which can be attributed to Si II absorptions formed in regions above the photosphere. Typical velocity of the HVF of Si II $\lambda$6355 is about $18,000$ km s$^{-1}$, while this HVF can reach above 30,000 km s$^{-1}$ for Ca II \citep{Mazzali05a,Mazzali05b,Tanaka08,Silverman15,Zhao15,Zhao16,Li20}. The HVF of Si II  $\lambda$6355 usually only appears in early spectra, while the HVF of Ca II HK or Ca II NIR may also appear in some post-maximum spectra \citep[][]{Hatano99, Mazzali05b,Childress14,Maguire14,Silverman15}. Our recent study revealed an anti-correlation between the properties of Si II $\lambda$6355 HVF and those of the O I $\lambda$7773 HVF \citep{Zhao16}, which suggests that the HVFs may be associated with IME burnings in the outermost layers \citep[see also, ][]{Kato18,Mulligan17,Mulligan19}.

The strength of absorption features can be quantified with the pseudo-equivalent width \citep[pEW; see, for example,][]{Hachinger06, Blondin11} or the absorption depth normalized to the pseudo-continuum \citep[see, for example,][]{Nugent95, Blondin11}. The excitation energy of the absorption feature is calculated by $E_{exc}=E_{upper}-E_{lower}=hc/\lambda$, where $E_{lower}$ is the lower energy level, and $E_{upper}$ is the upper energy level, $\lambda$ is the rest-wavelength\footnote{A table of the energy levels of important lines in spectra of SNe Ia near maximum light can be found on website: \url{http://supernova.lbl.gov/~dnkasen/tutorial/}}. The effect of $E_{lower}$ can be seen from the strong absorptions of PVFs (photospheric-velocity feature) and HVFs of Si II $\lambda$6355 and Ca II NIR \citep{Zhao15}, while the effect of $E_{exc}$ can be seen from comparison between lines Ca II HK and Ca II NIR, or between Si II $\lambda$4130 and Si II $\lambda$5972. The ionization degree also significantly affects the strength of absorption features, especially in 91T/99aa-like SNe \citep{Mazzali95,Fisher99,Hachinger08,Sasdelli14,Taubenberger17}.

While SNe Ia are known as standardizable candles, they actually have very diverse spectroscopic properties, including different velocity and strength of the observational features (including the HVFs), and presence/absence of some rare absorption features, such as variable sodium lines, H or He lines \citep[e.g.,][]{Patat07,Dilday12,Jacobson19}. The photospheric temperature, which is associated with the amount of $^{56}$Ni produced in the explosion \citep{Nugent95, Hoeflich96, Mazzali01, Kasen07,Churazov14}, is considered to be the main cause for the spectroscopic diversity, but even two SNe Ia with similar brightnesses could exhibit very different spectral features \citep[The opposite could also be the case; some SNe Ia with similar spectral features actually have significantly different peak luminosities; see, for example,][]{Foley20}. Physically, the spectroscopic property of SNe Ia is determined by the initial chemical compositions of the WD progenitors \citep[e.g.,][]{Hachinger06, Nomoto13}, the explosion and burning modes \citep[e.g.,][]{Whelan73, Nomoto82, Fink10, Maeda10, Bulla16, Maguire18}, the birthplace environments \citep[e.g.,][]{Wang13,Mandel17, Hill18,Meng19} and other factors. For example, the appearance of unburnt carbon feature C II $\lambda$6580 may indicate a large amount of carbon in the WD progenitor \citep{Parrent11, Thomas11, Folatelli12, Silverman12d, Hsiao15}, and tend to be related with the explosion mechanism/progenitor property \citep{Li20}. The near-ultraviolet (NUV) features of SNe Ia were found to be strongly correlated with the progenitor metallicity \citep{Ellis08, Mazzali14,Brown20}. The variable sodium or calcium feature reveals dense CSM around some SNe Ia \citep{Hamuy03,Patat07,Sternberg11,Maguire13,Wang19}, which is in favor of the single-degenerate scenario \citep{Whelan73,Nomoto82,Dilday12}. At the same time, there are also observational indications suggesting that the double-degenerate scenario may be responsible for some SNe Ia \citep[e.g.,][]{Li11,Schaefer12,Kerzendorf14}; clearly more study is necessary in this respect.

Spectroscopic information could be used to improve the accuracy of distance determinations of SNe Ia, either through spectroscopic parameters that indicate the intrinsic diversities \citep[]{Bailey09,Wang09,Foley11a,Maeda11,Blondin11,Silverman12c,Burns14,Mandel14,Brown18} or through others that reflect the explosion environments \citep[]{Wang13, Anderson15, Hill18}. For example, \citet{Wang09} found that the SN Ia subclass with high Si II $\lambda$6355 velocities ($v\ge11,800$ km s$^{-1}$, HV) near $B-$band maximum \footnote{Throughout this paper the maximum is referred to the $B-$band maximum.} and the subclass with normal Si II $\lambda$6355 velocities ($8,000<v<11,800$km s$^{-1}$, NV) near $B-$band maximum have different observed colors. Using different extinction ratios for the two subclasses reduced the luminosity dispersion from 0.178 mag to 0.125 mag. Also, there are some studies showing intrinsic correlation between the velocity and the light curve time scale \citep{Zhang10, Ganeshalingam11, Kawabata20}. Some characteristic features, i.e., C II $\lambda$6580 absorption, may help select a subsample of ``well-behaved" SNe Ia for more precise distance measurements \citep{Parrent11, Thomas11, Folatelli12, Silverman12d}.

Previous studies of the spectral features mostly focus on near-maximum light epochs \citep[e.g.,][]{Branch06, Hachinger06, Wang09, Blondin11, Silverman12b, Silverman12c}. Due to the lack of early-time spectra ($t\lessapprox-7$ d), study on early-time spectral behaviors of SN Ia is still limited. Some information, such as features from unburnt materials, are often seen in the early phase. Different models usually predict similar maximum spectra; the early spectral information can be a key to further constrain explosion models \citep[e.g.,][]{Maeda18}. A comparison between features of early time and near maximum light tells much about the influences of temperature, luminosity, and change in element abundances, which forms a fundamental problem for interpreting the spectral diversity of SN Ia \citep[e.g,][]{Blondin12, Folatelli13, Maguire14, Stahl20}. A comparison between different features also provides important clues about the physics behind the spectral diversity of SN Ia. In this work, we collect a large sample of early spectra of SNe Ia as well as spectra near maximum light, and carefully examine the evolutions of four important spectral features Si II $\lambda\lambda$4130, 5972, 6355 and S II W-trough. We also examine the differences and correlations between these features, and their dependences on the SN brightness. We address many details of the evolutions of these spectral features, and the results may be useful for future studies of SN Ia spectroscopy and SN la cosmology.

This paper is organized as follows. In Section \ref{S2} we present the
spectroscopic sample used in this study, and describe our measurement procedure. Section \ref{S3} presents the temporal evolutions of the velocities and pEWs for different subclasses, and explore the correlation between velocity and pEW. In Section \ref{S4} we focus on the link between spectral properties and the luminosity decline rate $\Delta$m$_{15}$($B$)\footnote{Defined as the change in $B$-band magnitude from peak to 15 days after that in the rest-frame of the SN (Phillips 1993).}. In Section \ref{S4} we also examine the temporal evolutions of the velocity-differences and depth-ratios between different features, and present the evolutions of correlations of several possible spectral luminosity indicators with $\Delta$m$_{15}$($B$). In Section \ref{S3} and \ref{S4}, the physical implications of our observational results are also discussed. We close the paper in Section \ref{S5} with our concluding remarks.

\section{Data and Measurements\label{S2}}

\emph{Data source}: Most spectra used in this work were compiled from the CfA supernova program \citep{Matheson08, Blondin12}, the Berkeley supernova program \citep{Silverman12a, Silverman12b, Stahl20} and the Carnegie Supernova Project \citep[CSP, ][]{Folatelli13}. These SNe Ia are mostly included in our previous study \citep{Zhao15}, where more information can be found. Photometric parameters are taken from the literature. After $t \approx$ +6 d\footnote{Throughout this paper the spectral phases correspond to the rest-frame days relative to the $B-$band maximum. All time intervals in this paper are also in rest-frame days.}, serious blends and spectral distortions make it difficult to perform an accurate measurement for most features of Si II and S II. Therefore our sample includes only spectra taken before +6 d. All spectra cover the wavelengths of the absorption features Si II $\lambda\lambda$5972, 6355 and S II W-trough (5,300 - 6,000 \AA).

\emph{Sample selection:} The original database contains more than 5,000 spectra of SNe Ia, but only about 2,000 spectra (from about 400 SNe Ia) were obtained before +6 d. We further selected about 120 SNe Ia which have at least 5 spectra at different phases between $-15$ and +5 d. The purpose of selecting relatively extensively observed SNe is to reduce the influence of the SN diversity on the mean evolutions. If using different objects at different phases, the evolutionary trends could be covered by the SN diversity, especially for those with slow evolutions. To take an extreme example, suppose we would use a sample including only two spectra from SN 1998aq at $-9.2$ d ($V_{Si6355}=10,944$ km s$^{-1}$) and SN 2003cg at $-6.2$ d ($V_{Si6355}=11,530$ km s$^{-1}$), a false-increasing trend of $V_{Si6355}$ appears. Also, for most objects, we used spectra from one source to avoid systematic error. This additional selection criteria further reduced the sample size to 105 SNe Ia. Furthermore, there were other problems such as poor spectra quality (i.e., low signal-to-noise ratio), too complex continuum, strong line contamination, insufficient wavelength coverage, or unsuitable phase distribution (for example, for SN 2000fa we have 5 spectra at $-11$, $-9$, +1, +2 and +4 days, but these data hardly reveal details of the evolutionary trend). By considering the above factors, we obtained our final sample with 79 SNe Ia.

\emph{The rest-wavelengths of the two components of S II W-trough:} Multiple lines contribute to the absorption feature of S II W-trough, merging together as a doublet absorption. The rest-wavelength of the bluer component is usually taken to be 5454 \AA. The rest-wavelength of the redder component, however, has been disputed, and most people use 5640 \AA~\citep[e.g.,][]{Hachinger06,Blondin11}. In this work, we also use 5454 \AA~as the rest-wavelength of the bluer component, but 5620 \AA~for the redder component. The adoption of 5620 \AA~rest-wavelength was calculated by (1.43219$\times$5606.15 \AA+1.40281$\times$5640.35 \AA)/(1.43219+1.40281)$\approx$5620 \AA, where 5606.15 and 5640.35 \AA~are the wavelengths of the two lines that blend into the redder component of S II W-trough, and 1.43219 and 1.40281 are the corresponding oscillator strengths, respectively. Using 5620 \AA~rest wavelength for the redder component leads to a mean velocity difference of about 400 km s$^{-1}$ between these two components, while using 5640 \AA~rest wavelength leads to a mean velocity difference of about 1,400 km s$^{-1}$. Since two lines with very similar energy levels are expected to have very similar velocities, it is more physically plausible to use 5620 \AA~as the rest wavelength of the redder component of S II W-trough ($E_{lower}$: 13.73 eV vs. 13.70 eV, $E_{exc}$: 2.21 eV vs. 2.27 eV).

\emph{Smoothing:} Before fitting, we smoothed each observed spectrum with a locally weighted linear regression \citep[nearer neighbors were more heavily weighted; see, for example,][]{Cleveland79}. For fittings of S II W-trough and Si II $\lambda\lambda$5972, 6355, the smoothing window was $125 - 150$ \AA. This span effectively reduces the influence of noise, and also well preserves the profiles of the features. For Si II $\lambda$4130, the span is much smaller ($\approx$35 \AA). This well preserves the details of Si II $\lambda$4130 which could be very weak at early times.

\begin{figure}\centering
\includegraphics[width=\columnwidth]{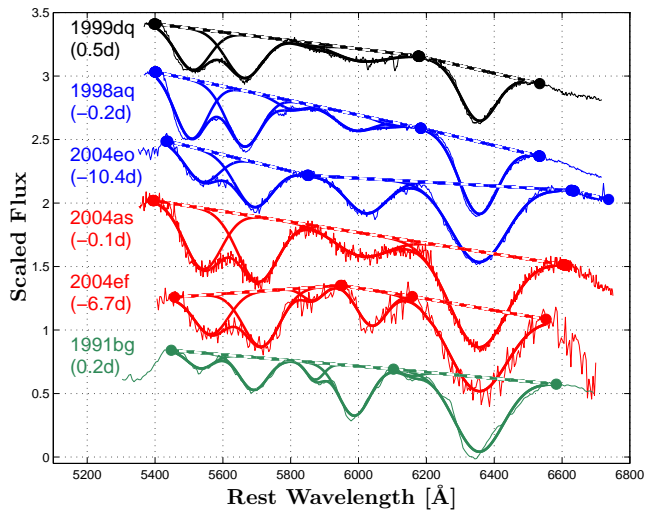}
\caption{Gaussian fits to the absorption features of Si II $\lambda\lambda$6355, 5972 and S II W-trough in the optical spectra of some representative SNe Ia near maximum light. Color coded by subclasses of \protect\citet[][]{Wang09} are spectra of NV (blue) SNe 1998aq and 2004eo, HV (red) SNe 2004as and 2004ef, 91T/99aa-like (black) SN 1999dq and 91bg-like (sea-green) SN 1991bg. The fluxes have been normalized, redshift-corrected and additive offsets applied for clarity, and labeled by phases with respective to $B-$band maximum light. The large dots mark the points that were considered to be featureless. The pseudo-continua were then constructed by connecting them with straight lines (marked with dash-lines). Si II $\lambda$4130 is truncated off the plot to focus on the region covered by more complex fitting.}
\label{Fig_fits} 
\end{figure}

\emph{Pseudo-continuum:}As demonstrated in Figure \ref{Fig_fits}, the pseudo-continua are defined by connecting the featureless points in the spectrum. Note that the two ends of a feature are not necessarily featureless, because there could be serious blending between two neighboring features. For example, the red end of Si II $\lambda$5972 of SN 2004as in Figure \ref{Fig_fits} is much lower than the straight line that well fit the featureless points near 5400, 5850 and 6600 \AA. The red end of S II $\lambda$5620 of SN 2004ef shows a similar behavior, as shown in Figure \ref{Fig_fits}. For some spectra at early times, more sophisticated method for determining the pseudo-continuum is required if a ``perfect" fitting is desired. This is however beyond the scope of this paper, and this does not affect the conclusions of this work. For most spectra, the pseudo-continuum is well-fitted by the classical method, i.e., connecting featureless points. In Figure \ref{Fig_fits}, typical examples of the pseudo-continuum are shown. In some cases the pseudo-continuum (between 5,400 and 6,500 \AA) appears to be linear, while in other cases the pseudo-continuum appears to be ``$\Lambda$"-shape (e.g., 2004eo at $-10.4$ d) or ``V"-shape (e.g., 2004ef at $-6.7$ d). 

\emph{Spectral fitting}: The measurement procedure is overall similar to that applied in our previous works \citep{Zhao15,Zhao16}. The Si II $\lambda$4130 feature was fitted with a single-gaussian model. S II W-trough, Si II $\lambda\lambda$5972 and 6355, HVF of Si II $\lambda$6355, C II $\lambda$6580 and an unknown feature near 5600 \AA~were fitted consistently and simultaneously with a 7-component gaussian function. In most cases, these features were well-separated by applying velocity constraints. For example, S II $\lambda$5454 with velocity 7,000 $\leq V_{S5454}\leq$ 12,000 km s$^{-1}$ is located at 5240 $\leq \lambda_{S5454}\leq$ 5328 \AA, while its neighboring feature S II $\lambda$5620 with similar velocity is located at 5400 $\leq \lambda_{S5620}\leq$ 5490 \AA. This means the two components of the S II W-trough will never be mistaken. All fittings were visually inspected to ensure that a good fit was obtained. The velocity and line strength measurements are listed in Table \ref{Tab1}.

\emph{Measurement uncertainties}: The smoothed-and-interpolated spectra (using ``linear-interpolation" method in MATLAB, with 0.5 \AA~interval) are generally well fitted by the multiple-gaussian function, with the coefficient of determination (R-squared) ranging from 0.925 to 0.999. However, for each component the uncertainty could be much higher due to the blendings. The uncertainties (for each component) are given in Table \ref{Tab1}. They are the total uncertainties with contributions from the gaussian fitting, the flux error, and the wavelength uncertainty (depending on the wavelength interval).

\begin{table*}
 \centering\scriptsize
\resizebox{\textwidth}{!}{
 \begin{threeparttable}
\caption{Measured velocities and line strengths. This is a sample of the full table, which is available online (see ``Supporting Information"). \label{Tab1}}

\begin{tabular}{ l  @{\extracolsep{\fill}}*{17}{c} }

\\\hline\hline\addlinespace
SN &Phase\tnote{a} &Subtype\tnote{b} &$\Delta$m$_{15}$($B$)\tnote{c} &$V_{Si}^{4130}$\tnote{d} &$W_{Si}^{4130}$\tnote{e} &$V_{Si}^{5972}$\tnote{f} &$W_{Si}^{5972}$\tnote{g} &$V_{Si}^{6355}$\tnote{h} &$W_{Si}^{6355}$\tnote{i} &$V_{S}^{5454}$\tnote{j} &$W_{S}^{5454}$\tnote{k} &$V_{S}^{5620}$\tnote{l} &$W_{S}^{5620}$\tnote{m} &Src\tnote{n} &Ref.\tnote{o}
\\\addlinespace
    &(d)  &  &(mag)  &(km s$^{-1}$) &(\AA)  &(km s$^{-1}$) &(\AA)  &(km s$^{-1}$)  &(\AA) &(km s$^{-1}$)  &(\AA)  &(km s$^{-1}$)  &(\AA) \\
\midrule

1994D & -11.5 & NV & 1.37 & 11447(573) & 10(1) & 12177(609) & 57(4) & 13677(684) & 139(3) & 10285(515) & 28(2) & 11022(552) & 44(2) & CfA & 1 \\
1994D & -9.5 & NV & 1.37 & 10321(517) & 14(2) & 11575(579) & 29(1) & 12395(620) & 102(3) & 10250(513) & 31(3) & 10924(547) & 42(2) & CfA & 1 \\
1994D & -8.5 & NV & 1.37 & 10597(530) & 19(2) & 11370(569) & 23(2) & 12169(609) & 105(3) & 10398(520) & 27(2) & 11083(555) & 37(2) & CfA & 1 \\
1994D & -4.5 & NV & 1.37 & 10351(518) & 19(2) & 10847(543) & 18(2) & 11584(580) & 95(7) & 9610(481) & 31(3) & 10153(508) & 35(3) & CfA & 1 \\
1994D & -2.5 & NV & 1.37 & 10067(504) & 23(1) & 10603(531) & 19(2) & 11054(553) & 97(7) & 9362(469) & 38(3) & 9852(493) & 39(3) & CfA & 1 \\
1994D & 1.5 & NV & 1.37 & 9962(499) & 23(1) & 10782(540) & 24(2) & 10888(545) & 104(8) & 8770(439) & 40(3) & 9322(467) & 45(2) & CfA & 1 \\
1994D & 2.5 & NV & 1.37 & -- & -- & 11197(560) & 31(3) & 10813(541) & 106(8) & 8544(428) & 42(4) & 9252(463) & 54(2) & CfA & 1 \\
1994D & 3.5 & NV & 1.37 & 9146(458) & 20(1) & 11246(563) & 34(3) & 10746(538) & 109(9) & 8353(418) & 42(4) & 8824(442) & 56(2) & CfA & 1 \\
1994D & 4.5 & NV & 1.37 & -- & -- & 11293(565) & 32(2) & 10793(540) & 106(8) & 8361(419) & 38(3) & 8788(440) & 51(2) & CfA & 1 \\

\bottomrule
\end{tabular}
\scriptsize
\textbf{Notes.} Uncertainties are listed in parentheses.
\begin{tablenotes}
\item[a] {Phases of spectra are in rest-frame days relative to the time of $B-$band maximum light.}
\item[b] {Wang-scheme classification \citep{Wang09}. The velocity boundary between HV-SNe and NV-SNe is around 11,800 km s$^{-1}$.}
\item[c] {$B-$band light-curve decline rate $\Delta$m$_{15}$($B$).}
\item[d] {Velocity of Si~II $\lambda$4130, for the photospheric (PHO) component;}
\item[e] {Pseudo-equivalent width of Si~II $\lambda$4130, for the PHO component;}
\item[f] {Velocity of Si~II $\lambda$5972, for the PHO component;}
\item[g] {Pseudo-equivalent width of Si~II $\lambda$5972, for the PHO component;}
\item[h] {Velocity of Si~II $\lambda$6355, for the PHO component;}
\item[i] {Pseudo-equivalent width of Si~II $\lambda$6355, for the PHO component;}
\item[j] {Velocity of S~II $\lambda$5454, for the PHO component;}
\item[k] {Pseudo-equivalent width of S~II $\lambda$5454, for the PHO component;}
\item[l] {Velocity of S~II $\lambda$5620, for the PHO component;}
\item[m] {Pseudo-equivalent width of S~II $\lambda$5620, for the PHO component;}
\item[n] {Data source References: CfA- CfA supernova program \citep{Matheson08,Blondin12}; Lick- Berkeley Supernova Program \citep{Silverman12a,Silverman12b,Stahl20}; CSP- Carnegie Supernova Project \citep[CSP][]{Folatelli13}; THU- Tsinghua Supernova Program \citep{Zhang15a}; Alt07 = \citep{Altavilla07}; Chi13 = \citep{Childress13}; Per13 = \citep{Pereira13}; Zha15 = \citep{Zhang15b}; Zhe13 = \citep{Zheng13}}. Most of the spectra are available on the online ``open supernova catalog" \citep{Guillochon17};
\item[o] {Photometric References: 1 = \citep{Blondin12}; 2 = \citep{Brown19}; 3 = \citep{Childress13}; 4 = \citep{Childress14}; 5 = \citep{Folatelli12}; 6 = \citep{Folatelli13}; 7 = \citep{Ganeshalingam10}; 8 = \citep{Jha06}; 9 = \citep{Pignata04}; 10 = \citep{Silverman15}; 11 = \citep{Stritzinger11}; 12 = \citep{Wang09}; 13 = \citep{Zhang15b}; 14 = \citep{Zheng17};}

\end{tablenotes}
\end{threeparttable}}
\end{table*}

\section{Time Evolution of The Si II and S II Features\label{S3}}

\subsection{Temporal evolution of line velocity\label{S.vt}}
The velocity of a given feature is derived from the blueshift of its deepest absorption (i.e., the minimum of the feature after the pseudo-continuum is removed) relative to the rest-wavelength. Each feature in the spectrum has its own velocity, depending on where the absorption is located in the ejecta \citep[e.g.,][]{Patat96}. The velocity evolution of Si II $\lambda$6355 has been studied intensively \citep[e.g.,][]{Benetti05, Wang09, Foley11b, Blondin12}, while the velocity evolution for other lines may need further investigation. Velocity evolutions of lines Si II $\lambda\lambda$4130, 5972 and S II W-trough have been studied by, e.g., \citet{Silverman12b}, \citet{Folatelli13}, and \citet{Stahl20}. Our sample is most similar to that of \citet{Blondin12}, who did not give these velocity evolution measurements. Below we present the velocity evolutions for these lines as well as Si II $\lambda$6355.

\begin{figure*}\centering
\includegraphics[height=80mm,width=130mm]{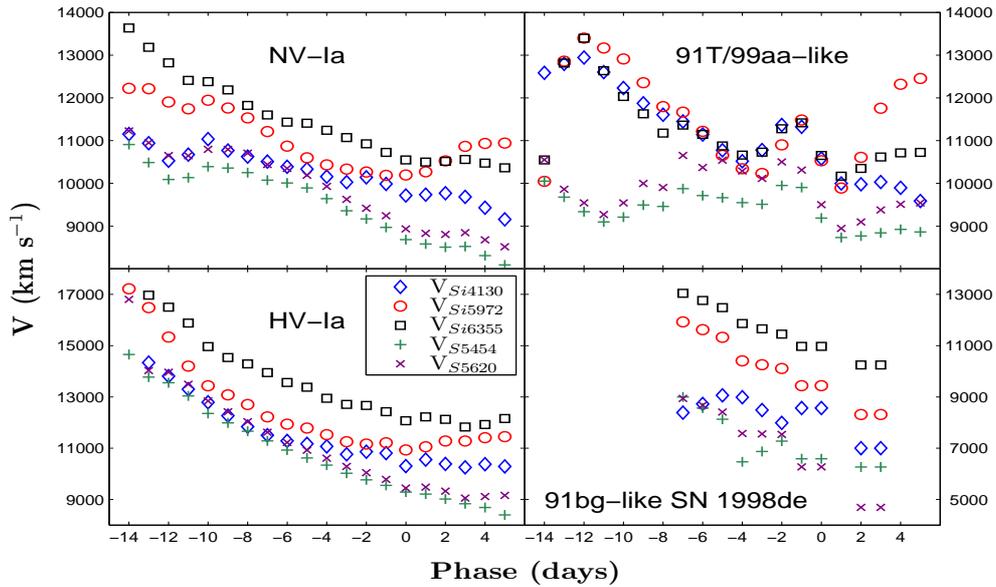}
\caption{Evolutions of the velocities (V) with time for the four subclasses defined by \protect\cite{Wang09}. The velocities are averaged over 1 d intervals. Color coded are the four lines we study: Si II $\lambda$4130 (blue diamond); Si II $\lambda$5972 (red circle); Si II $\lambda$6355 (black square); the two components of S II W-trough (green/brown). Due to lack of extensively observed 91bg-like objects, we can only use SN 1998de as a representative. Note that the y-axis ranges have been increased in the bottom panels to encompass the larger variations in the velocities of HV and 91bg-like SNe.}\label{Fig_tevo_V}
\end{figure*}

Figure \ref{Fig_tevo_V} shows the velocity evolutions for different subclasses in the \citet{Wang09} classification scheme. For each point, we took the mean of the measurements within $\pm0.5$ d, i.e, $t\pm0.5$ d. In general, the velocities decrease with time as expected (due to the recession of the photosphere following the expansion and the decrease in density). The only exception is the velocity of Si II $\lambda$5972 ($V_{Si5972}$), which somehow surprisingly increases with time after $t \approx$ $-2$ d, and even surpasses the velocity of Si II $\lambda$6355 ($V_{Si6355}$) after $t \approx$ +2 d (see discussion in \S \ref{S.cmp}). No systematic bias was found in the analysis method that could cause the abnormal behavior of $V_{Si5972}$.

During the phase from $t \approx$ $-14$ to +5 d, the NV objects have an average velocity gradient of about 200 km s$^{-1}$d$^{-1}$, while the HV objects have corresponding value of about 300 km s$^{-1}$d$^{-1}$, and the 91T/99aa-like objects \citep[][]{Filippenko92b,Phillips92,Li01,Garnavich04,Sasdelli14,Taubenberger17} have an average velocity gradient of about 100 km s$^{-1}$d$^{-1}$. During the phase from $t \approx$ $-7$ to +3 d, the representative 91bg-like \citep[][]{Filippenko92a, Mazzali97, Howell01,Hachinger09,Taubenberger17} object, SN 1998de is found to have a velocity gradient of about 400 km s$^{-1}$d$^{-1}$. It is clear that faster decliners tend to have more rapidly declining velocities, possibly due to their faster decreases in temperature.

Velocities obtained here are generally consistent with previous studies, including the rising trend of $V_{Si5972}$ after $B-$band maximum \citep[e.g.,][]{Silverman12b}. We use a sample that is most similar to \citet{Blondin12}. The $V_{Si6355}$ measured around the $B-$band maximum is about 11,000 \kms~for normal SNe Ia in \citet{Blondin12}, compared to about 10,600 \kms~obtained in our work. Minor difference could be due to two reasons: Firstly, the determination of the position of the absorption minima. Whether it is directly determined from the smoothed flux or determined by a gaussian fitting can lead to different results in the velocity measurement. Secondly, the blending of some features (e.g., Si II $\lambda$6355 at early times) with other lines or the HVF may also seriously affect the velocity measurement \citep[e.g.,][]{Zhao15}.

\subsection{Temporal evolution of line strength\label{S.wt}}

The absorption strength can be quantified by the pseudo-equivalent width \citep[pEW, see, e.g.,][ for details]{Blondin11, Zhao15}, or the absorption depth ($H$) which is defined as the deepest absorption of the feature normalized to the pseudo-continuum. The temporal evolutions of the pEWs of Si II $\lambda\lambda$4130, 5972, 6355 and S II W-trough have been studied by, e.g., \citet{Silverman12b}, \citet{Folatelli13}, and \citet{Stahl20}, but not given by \citet{Blondin12}. As mentioned previously, our sample is very similar to that of \citet{Blondin12}, and below we present the pEW evolutions for our sample.

\begin{figure*}\centering
\includegraphics[height=80mm,width=130mm]{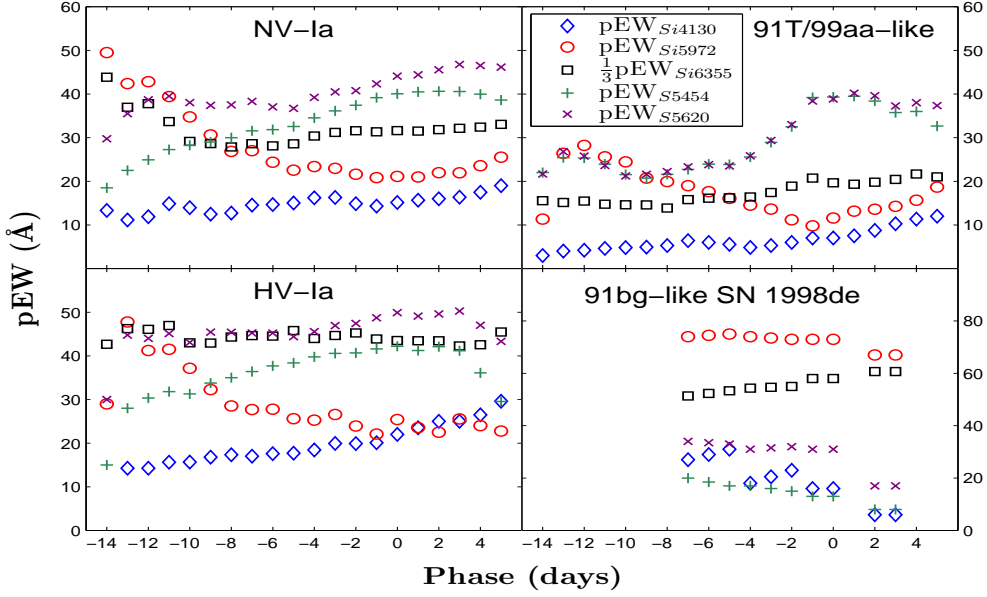}
\caption{Evolutions of the pseudo-equivalent widths (pEWs) with time for the four spectroscopic subclasses defined by \protect\cite{Wang09}. The pEWs are averaged over 1 d intervals. Due to lack of extensively observed 91bg-like object, we can only use SN 1998de as a representative. Colors and shapes of data
points are the same as in Figure \ref{Fig_tevo_V}.}
\label{Fig_tevo_EW}
\end{figure*}

Figure \ref{Fig_tevo_EW} shows the pEW evolutions for different subclasses in the classification scheme proposed by \citet{Wang09}. Each point was averaged over 1 d intervals, i.e, $t\pm0.5$ d. \emph{NV objects:} In general, the pEWs of NV objects increase with time before the $B-$band maximum. An exception for this tendency is pEW of Si II $\lambda$5972, which decreases until $t \approx$ $-1$ d. The other longer-wavelength feature, Si II $\lambda$6355, also weakens with time before $t \approx$ $-7$ d, then slowly increases until $t \approx$ $-4$ d, after which it remains virtually unchanged. It appears from Figure \ref{Fig_tevo_EW} that the pEW of S II W-trough of NV objects reaches a maximum value near $t \approx$ +2 d, though the trend is less noticeable than in 1991T/99aa-like SNe Ia. For Si II $\lambda$4130, though the average pEW shows a slight trend of increasing after +2 d, there are actually more than half of NV objects exhibiting a decrease or plateau of pEW after $t \approx$ +2 d. Specifically, after +2 d, SNe 1994D, 2002er, 2003cg and 2005cf show a decreasing pEW, SNe 1998aq, 2004at, 2005el, 2006le and 2011fe show constant pEW, while SNe 2002de, 2003du, 2006S, 2007af, 2008ar, 2008bf and 2009ab show a slowly increasing pEW (see Table \ref{Tab1}).

The delay in reaching the maximum line-strength relative to the maximum light could be due to our using $B-$band maximum light as the 0 d, as $V-$band maximum is usually 1 - 2 d later than the $B-$band maximum. We notice that Si II $\lambda$5972 behaves differently than other features, with a minimal pEW near $t \approx$ $-1$ d (see discussion in \S \ref{S.cmp}). The fact that the features generally reach an extremum (i.e., maximum or minimum) pEW near maximum light suggests that the line strengths are significantly influenced by the luminosity. However some features (i.e., Si II $\lambda$6355) seem not to follow the trend. This can be explained as other factors such as saturation, temperature, element abundance, which can also affect the absorption strengths.

\emph{Other subtypes:} HV objects are characterized by large and nearly constant line strengths of Si II $\lambda$6355 and S II W-trough, which may imply a high degree of saturation. The evolutionary trends of the velocities and the pEWs of HV SNe are overall similar to those of NV SNe, supporting the classification of both ``HV-Ia" and ``NV-Ia" objects as ``normal" SNe Ia. The 91T/99aa-like objects are characterized by weak Si II $\lambda$6355 and nearly constant strength of Si II $\lambda$4130.

The evolutionary trends we showed above are generally consistent with the results given by some recent studies \citep{Silverman12b, Folatelli13, Stahl20}, except for $pEW_{Si5972}$. Our results are similar to that of \citet{Folatelli13}, with a decreasing trend of $pEW_{Si5972}$ before $B-$band maximum. In \citet{Stahl20}, four-day mean value of $pEW_{Si5972}$ slightly decreases from 20 \AA~to 17 \AA~at $-17<t<-9$ d, but it then increases progressively from 17 to 25 \AA~at $-9<t<+5$ d. \citet{Stahl20} also gives the four-day mean value of $pEW_{Si5972}$ using the data from \citet{Silverman12b}, which increases from 18 to 26 \AA~at $-12<t<-5$ d, it then remains nearly unchanged between $-5$ and +5 d ($\approx$ 26 \AA). The discrepancy may be due to the different sample or different measurement method (see discussion in \S \ref{S2}). Nevertheless, our measurement of $pEW_{Si5972}$ is still roughly consistent with the previous studies. For example, at $B-$band maximum, \citet{Silverman12b} give a mean value of $pEW_{Si5972} \approx 25$ \AA~for normal SNe, while we give a corresponding value of $\approx 23$ \AA. At $-5$ d, \citet{Silverman12b} give a mean value of $\approx 26$ \AA~for normal SNe, while we give $pEW_{Si5972} \approx 25$ \AA.

\subsection{Correlation between velocity and line strength\label{S.vw}}

The correlation between $V_{Si6355}$ and $pEW_{Si6355}$ has been previously studied by, e.g., \citet{Wang09}, \citet{Blondin12} and \citet{Maguire14}. In \citet{Wang09}, a clear trend of larger pEW corresponding to higher velocity at maximum light was observed in both NV and HV subclasses. In \citet{Blondin12}, HV SNe show positive velocity-pEW correlation, while NV SNe show no clear velocity-pEW correlation. In \citet{Maguire14}, HV SNe show no clear velocity-pEW correlation, while NV SNe show a slight trend with larger pEW corresponding to larger velocity. The discrepancy between these results may be due to sample selection. Also note that the correlation could be seriously affected by the evolutionary effect, depending on how velocity and pEW evolve in the phase range. In this section, we do not reinvestigate the velocity-pEW correlation at maximum light, but instead focus our analysis on the evolution of this correlation.

\begin{figure*}\centering
\includegraphics[height=100mm,width=130mm]{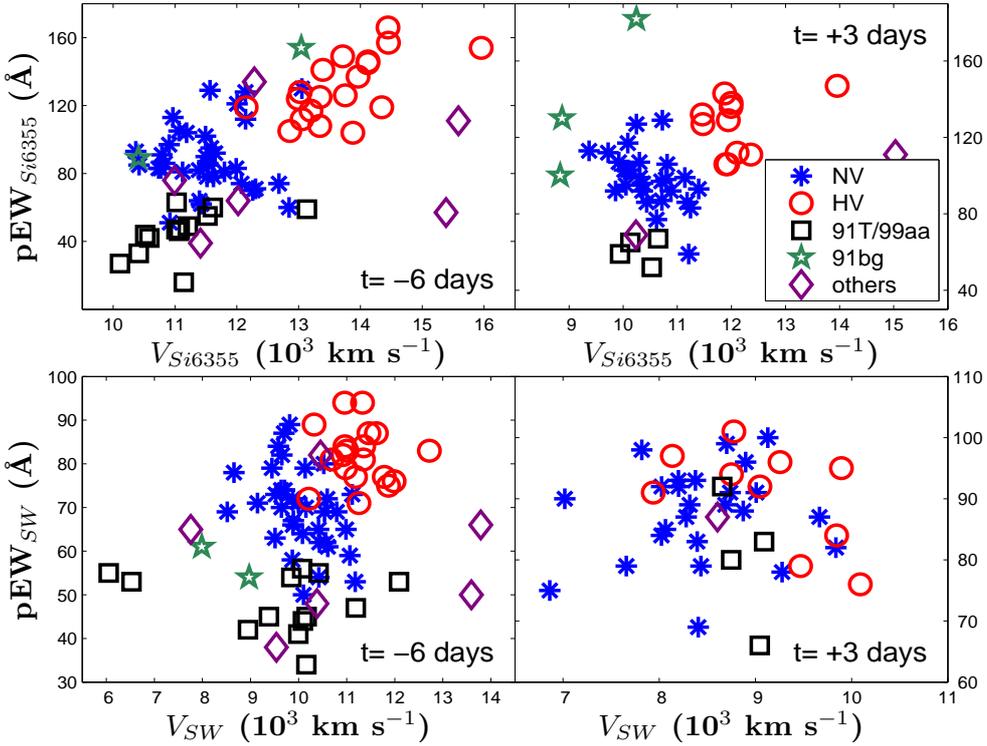} \caption{ Top-left panel: pEW of the Si II $\lambda$6355 feature as a function of the velocity of Si II $\lambda$6355 feature for SN Ia spectra within 2 d of $-6$ d since maximum light. If more than one measurement was available, the one closest to the central phases was used. Bottom-left panel: Similar to the top-left panel, but for the S II W-trough feature. Right panels: Similar to the left panels, but for spectra near +3 d. The sample has been split into subclasses as defined by \protect\citet{Wang09}. Blue asterisks are normal-velocity (NV) objects, red circles are high-velocity (HV) objects, black squares are 91T/99aa-like objects, green pentagrams are 91bg-like objects. Other objects, including peculiar or uncertain subtype objects are marked with brown diamond.}
\label{Fig_V_vs_W}
\end{figure*}

Figure \ref{Fig_V_vs_W} shows the correlation between velocity and pEW for Si II $\lambda$6355 and S II W-trough. The data are restricted to be within $\pm2$ d from the selected epochs to reduce the evolutionary effect: left panels are for the sample at $-6$ $\pm 2$ d, right panels of the same figure are for the sample at 3 $ \pm 2$ d. The overall trend is positive, but mainly due to the fact that HV SNe generally have both higher velocities and pEWs than the NV counterparts. Within each subclass, the exact correlation seems to depend on the phase.

For Si II $\lambda$6355 in the spectra of HV or 91T/99aa-like SNe, the velocity and pEW appear to be positively correlated near $t \approx$ $-6$ d, but this correlation becomes weak near +3 d. The reason of the positive velocity-pEW correlation near $t \approx$ $-6$ d is unclear. Physically, the velocity and pEW may not affect each other directly, but they might be indirectly connected through temperature, luminosity or element abundance. For the NV SNe, the velocity-pEW correlation of Si II $\lambda$6355 is weak if there is any near $-6$ d, and seems to show a negative correlation near $t \approx$ +3 d. A negative correlation can be explained as a result of lower optical depth in outer layers, in addition to the relatively high degree of homogeneity within NV-Ia subclass, which reduces the effect of element abundance and luminosity. Negative velocity-pEW correlation is also seen in both photospheric and HVF components of O I $\lambda$7773 \citep{Zhao16}. The change in $V_{Si6355}-pEW_{Si6355}$ may be related to the changes in $\Delta$m$_{15}$($B$)-dependence of $V_{Si6355}$ and $pEW_{Si6355}$. At $t \approx$ +3 d, possibly due to decrease in temperature, $V_{Si6355}$ becomes anti-correlated with $\Delta$m$_{15}$($B$), while $pEW_{Si6355}$ is positively correlated with $\Delta$m$_{15}$($B$) (see \S \ref{S.vdm} and \ref{S.wdm} for more details). 

For S II W-trough near $t \approx$ $-6$ d, the velocity-pEW correlation appears to be insignificant within the HV or 91T/99aa-like objects, and it may be negative for the NV objects. This negative correlation for NV objects somehow disappears near $t \approx$ +3 d, as shown in the bottom-right panel of Figure \ref{Fig_V_vs_W}. This may be due to a negative correlation between $V_{SW}$ and $\Delta$m$_{15}$($B$), which is weak at $t \approx$ $-6$ d, but strong at $t \approx$ +3 d (discussed later in \S \ref{S.vdm} and \ref{S.wdm}).

\subsection{Comparison between 91T/99aa-like and 91bg-like SNe\label{S.91T91bg}}

The effects of the temperature on the absorption features have been introduced in \S \ref{S1}. Here, we make comparison between the ``hot" 91T/99aa-like objects and the ``cool" 91bg-like objects to see the details. (a) As shown in Figure \ref{Fig_tevo_V}, the velocities of 91T/99aa-like objects stay almost unchanged after $t \approx$ $-8$ d relative to $B-$band maximum. In contrast, the velocities of 91bg-like objects decrease rapidly with time; (b) As shown in Figure \ref{Fig_tevo_EW}, the pEWs of 91T/99aa-like objects keep increasing with time. On the other hand, the pEWs of 91bg-like objects keep decreasing with time, except for the lowly excited feature Si II $\lambda$6355. (c) As shown in Figure \ref{Fig_V_vs_W} (which shows relations between the velocities and pEW or Si II $\lambda$6355 and S II W-trough at different epochs), 91T/99aa-like objects have much weaker Si II $\lambda$6355 but much stronger S II W-trough than 91bg-like objects. These may serve as an example of temperature effect, which is important for understanding the spectral diversity, as will be further discussed in the next section.

\section{The decline-rate dependences of the Si II and S II features\label{S4}}

\subsection{The decline-rate dependence of the velocity\label{S.vdm}}

While some studies \citep{Hatano00, Benetti05, Foley11b} suggested no clear correlation between the velocity of Si II $\lambda$6355 at $B-$band maximum and $\Delta$m$_{15}$($B$), others suggest that the velocity at $B-$band maximum is anti-correlated with $\Delta$m$_{15}$($B$) \citep{Hachinger06}. In this section, we show how the velocity-$\Delta$m$_{15}$($B$) correlation evolves with time.

\begin{figure}\centering
\includegraphics[width=\columnwidth]{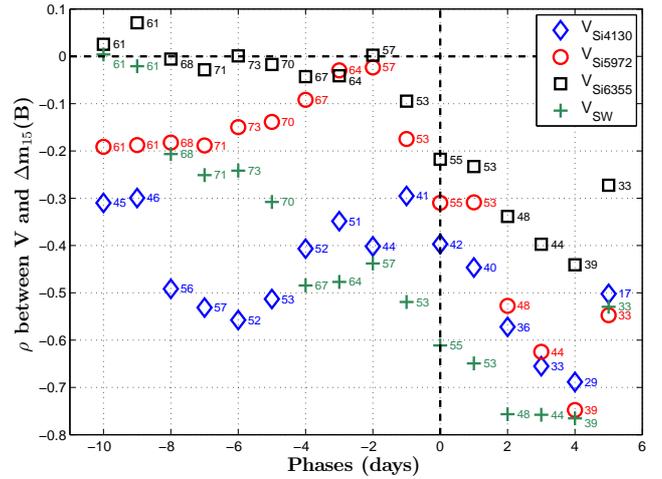}
\caption{Time-evolutions of the Pearson correlation coefficients between the velocities and $\Delta$m$_{15}$($B$). Measurements within 2 d were used, and if more than one measurement was available, the one closest to the central phases was used. The whole sample, including NV, HV, 91T/99aa-like and 91bg-like SNe are all included.}
\label{Fig_tevo_corr_v}
\end{figure}

Figure \ref{Fig_tevo_corr_v} shows the Pearson correlation coefficients ($\rho$, as defined in Eq. \ref{eq1}) between the velocities and the $\Delta$m$_{15}$($B$). The negative Pearson coefficients suggest that the velocities are generally anti-correlated with the $\Delta$m$_{15}$($B$). The Pearson coefficients are mostly very small, suggesting that the correlations are modest, insignificant or even nonexistent at most phases. The Pearson correlation coefficient is defined as follows:

\begin{equation}\label{eq1}
  \rho_{x,y}=\frac{cov(x,y)}{\sigma_x\sigma_y}=\frac{\sum^{n}_{i=1}(x_i-\bar{x})(y_i-\bar{y})}{\sqrt{\sum^{n}_{i=1}(x_i-\bar{x})^2}\sum^{n}_{i=1}(y_i-\bar{y})^2},
\end{equation}
where $\sigma_x$, $\sigma_y$ are the standard deviations of $x$ and $y$, respectively, and $cov(x,y)$ is the covariance between $x$ and $y$.

\begin{figure}\centering
\includegraphics[width=\columnwidth]{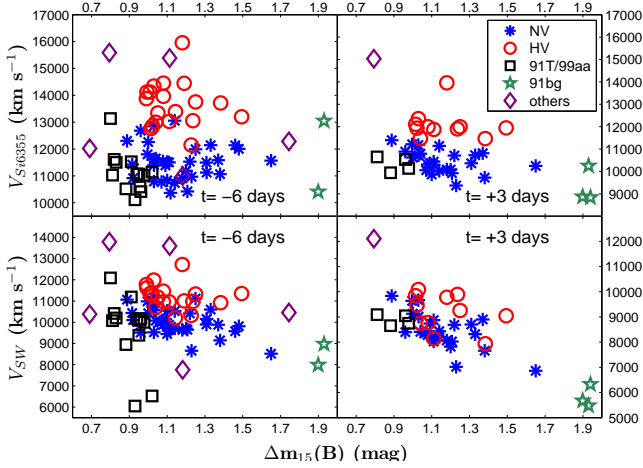} \caption{Velocities of Si II $\lambda$6355 and S II W-trough at $-6$ and +3 d as a function of $\Delta$m$_{15}$($B$). Measurements within 2 d were used, and if more than one measurement was available, the one closest to the central phase was used. Colors and shapes of data points are the same as in Figure \ref{Fig_V_vs_W}.}
\label{Fig_dm_vs_V6355SW}
\end{figure}

Figure \ref{Fig_tevo_corr_v} also shows how the correlations evolve with time. The correlations are very weak at early times, they then grow stronger with time until +4 d, implying a stronger dependence of the velocity on $\Delta$m$_{15}$($B$) at cooler temperature. For some reason the correlations between the velocities and $\Delta$m$_{15}$($B$) weaken with time after +4 d. The correlation is strongest for S II W-trough, and weakest for Si II $\lambda$6355, implying a stronger dependence on the SN brightness for lines from highly-excited levels. A relatively strong correlation seems to exist between $V_{SW}$ at $t \approx$ +3 d and $\Delta$m$_{15}$($B$), as shown in the bottom-right panel of Figure \ref{Fig_dm_vs_V6355SW}.

When the sample is restricted to only a subclass, the correlation of the velocity with $\Delta$m$_{15}$($B$) appears somewhat weaker. In fact, as shown in Figure \ref{Fig_dm_vs_V6355SW}, the correlation is largely driven by 91bg-like objects which have both lower velocities and luminosities than other subtypes. This may explain the relatively strong velocity-$\Delta$m$_{15}$($B$) correlation in \citet{Hachinger06}, as their sample contains a higher fraction of SN 91bg-like
events.

At $-6$ d, line velocities of 91T/99aa-like SNe seem to have stronger correlations with $\Delta$m$_{15}$($B$) than that of other subclasses, as shown in Figure \ref{Fig_dm_vs_V6355SW}. This may be due to a more significant ionization effect in 91T/99aa-like object, as their photospheric temperatures are relatively high \citep{Benetti05}. A higher degree of ionization can cause the singly ionized absorptions to move outward to the cooler, faster-moving layers, and thereby an anti-correlation between the velocity and $\Delta$m$_{15}$($B$). It is also noted that $V_{SW}$ appears to have a stronger correlation with $\Delta$m$_{15}$($B$) than $V_{Si6355}$. For example, at $-6$ d, when $\Delta$m$_{15}$($B$) increases from 0.8 to 1.0 mag, $V_{SW}$ (of 91T/99aa-like SNe) decreases from 12,000 to 6,000 km s$^{-1}$, while $V_{Si6355}$ (of 91T/99aa-like SNe) only decreases from 13,000 to 10,000 km s$^{-1}$, as shown in the left panels of Figure \ref{Fig_dm_vs_V6355SW} (see also the discussion in \S \ref{S.dv}). A possible reason is the relatively high energy levels of S II W-trough, which makes the line more sensitive to the photospheric temperature.

\subsection{The decline-rate dependence of the line strength\label{S.wdm}}

\citet{Hachinger06} and \citet{Blondin12} have studied the correlation between pEW and $\Delta$m$_{15}$($B$) for various lines at maximum light, with the aim to find better spectroscopic luminosity indicator. In this section, we further explore the temporal evolutions of the correlations between the line strengths and the $\Delta$m$_{15}$($B$), and compare them among different subclasses of SNe Ia. Unlike most studies, here we use the absorption depth ($H$) to represent the line strength, as it shows a slightly stronger correlation with $\Delta$m$_{15}$($B$) (see \S \ref{S.dmind} for more details). 

\begin{figure}\centering
\includegraphics[width=\columnwidth]{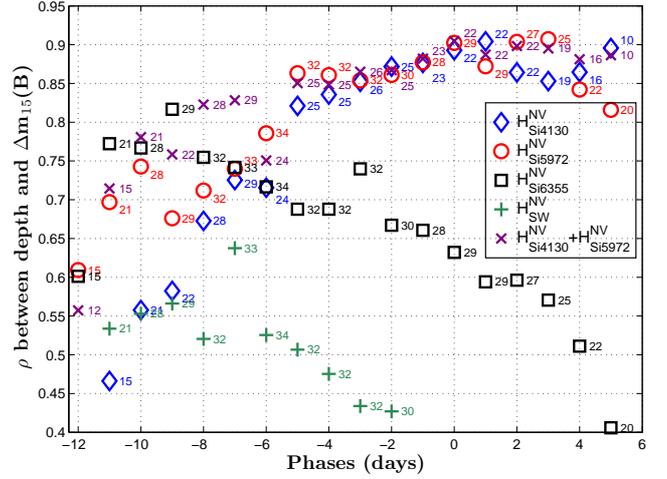}
\caption{Time-evolutions of the correlations between several indicators and the $\Delta$m$_{15}$($B$). Superscript ``NV" denotes the NV subclass which has a stronger correlation with $\Delta$m$_{15}$($B$) than other subclasses. Measurements within 2 d were used, and if more than one measurement was available, the one closest to the central phases was used. The sample size is marked on the right for each point.}
\label{Fig_dmInd}
\end{figure}

Figure \ref{Fig_dmInd} shows positive correlations between the depths of Si II and S II absorption features and $\Delta$m$_{15}$($B$). The sample is restricted to NV-SNe Ia whose high degree of homogeneity (meaning relatively similar velocities and line strengths of IME features, as well as relatively similar luminosities) leads to a tighter relation with $\Delta$m$_{15}$($B$). One possible explanation for the positive correlations is that the dimmer objects tend to suffer less burning and have hence more IME left in the ejecta. Another possible reason is that in dimmer objects the photospheric temperature is relatively low, in favor of low-ionization lines such as Si II $\lambda$6355.

It is somewhat surprising that dimmer objects display stronger absorptions in Si II $\lambda\lambda$4130, 5972 and S II W-trough than brighter objects, since these lines originate from rather highly excited levels. These positive correlations between the line strengths and $\Delta$m$_{15}$($B$), together with the pEW-evolutions (see \S \ref{S.wt}), seem to suggest a less importance of temperature effect on those prominent features than the luminosity. Nonetheless, the correlation is weak for Si II $\lambda$6355 which has the lowest excitation energy, and also weak for S II W-trough which has the highest excitation energy, suggesting that only the lines with proper (excitation \& ionization) energies may have a strong correlation with the peak brightness.

\begin{figure}\centering
\includegraphics[width=\columnwidth]{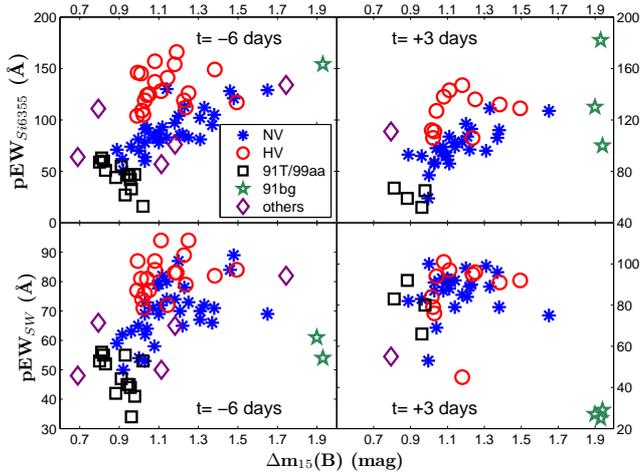}
\caption{A comparison of the decline-rate dependences of the pEWs of Si II $\lambda$6355 and S II W-trough at $-6$ d (left panels) and +3 d (right panels). Upper panels: pEWs of Si II $\lambda$6355 as a function of the decline rate. Lower panels: pEWs of S II W-trough as a function of the decline rate. Measurements within 2 d were used, and if more than one measurement was available, the one closest to the central phases was used. Colors and shapes of data points are the same as in Figure \ref{Fig_V_vs_W}.}
\label{Fig_dm_vs_W6355SW}
\end{figure}

In the left panels of Figure \ref{Fig_dm_vs_W6355SW}, we compare the decline-rate dependence of $pEW_{Si6355}$ at $t \approx$ $-6$ d to that of $pEW_{SW}$. Although these two lines are produced with very different excitation and ionization energies, it appears that their pEWs have similar correlations with $\Delta$m$_{15}$($B$). For NV SNe, the pEWs appear to be positively correlated with $\Delta$m$_{15}$($B$), which has been discussed in the previous paragraph. For HV SNe which have relatively large pEWs, the pEW-$\Delta$m$_{15}$($B$) correlation is insignificant.

In 91T/99aa-like SNe, as shown in Figure \ref{Fig_dm_vs_W6355SW}, the pEWs of both Si II and S II lines appear anti-correlated with the decline rate, i.e., brighter 91T/99aa-like objects tend to have stronger absorptions of Si II $\lambda$6355 and S II W-trough. This is rather puzzling because brighter 91T/99aa-like objects are expected to have more complete burning of IME to $^{56}$Ni. It may not be a result of either temperature effect or luminosity effect, because a dimmer/colder condition prefers the formation of Si II $\lambda$6355. A possible explanation is that oxygen burning in 91T/99aa-like object may contribute substantially to the formation of IME. Brighter 91T/99aa-like SNe may then have more IME than the dimmer ones. Moreover, for 91T/99aa-like objects, $pEW_{Si6355}$ appears to have a stronger correlation with $\Delta$m$_{15}$($B$) than $pEW_{SW}$. For example, at $-6$ d, when $\Delta$m$_{15}$($B$) increases from 0.8 to 1.0 mag, $pEW_{Si6355}$ decreases from 70 to 10 \AA, while $pEW_{SW}$ only decreases from 60 to 30 \AA, as shown in the left panels of Figure \ref{Fig_dm_vs_W6355SW}. This may be due to the relatively low abundance of Si II in 91T/99aa-like objects. 

\emph{Time evolution of the correlation between line strength and $\Delta$m$_{15}$($B$):} As can be seen from Figure \ref{Fig_dmInd}, the correlations of $H_{Si4130}$ and $H_{Si5972}$ with $\Delta$m$_{15}$($B$) tend to get stronger from about $-10$ to about +2 d. This may be due to the increase of line strength (meaning more absorptions of the photons and thus a tighter relation with the luminosity), less blending with HVF or neighboring features, and better determination of the pseudo-continuum (see \S \ref{S2}). The correlations of $H_{Si6355}$ and $H_{SW}$ with $\Delta$m$_{15}$($B$), on the contrary, tend to become weaker with time during the same period. The weakening of $H_{Si6355}$ correlation with $\Delta$m$_{15}$($B$) may be caused by line saturating, as a result of luminosity increase. As one can see from Figure \ref{Fig_dm_vs_W6355SW}, the pEW of Si II $\lambda$6355 at $t \approx$ +3 d barely increases after $\Delta$m$_{15}$($B$)$=1.4$ mag. The weakening of $H_{SW}$ correlation with $\Delta$m$_{15}$($B$) may be due to the decrease of temperature. As shown in the bottom-right panel of Figure \ref{Fig_dm_vs_W6355SW}, at $t \approx$ +3 d, $pEW_{SW}$ decreases sharply after $\Delta$m$_{15}$($B$)$=1.4$ mag, which could be a sign of too low temperature condition for the highly excited line S II W-trough.

\subsection{Velocity differences\label{S.dv}}

The ejecta velocity of SNe Ia varies from object to object, partially due to the asymmetric explosion and line-of-sight effect \citep{Maeda10,Maeda11}, and partially due to SN Ia diversity that arises from different progenitor systems and/or explosion mechanisms (see the references given in \S \ref{S1}). Even in the same spectrum, different lines still show different velocities, suggesting a layered structure of the SN Ia ejecta (with velocity increasing outward). In this section, we explore the temporal evolution of the velocity differences shown between the four spectral features that we study here, and their correlations with $\Delta$m$_{15}$($B$). Unless otherwise stated, throughout this work a velocity-difference between two features of the same species is defined as the velocity of the longer-wavelength feature minus the velocity of the shorter-wavelength feature.

\begin{figure}\centering
\includegraphics[width=\columnwidth]{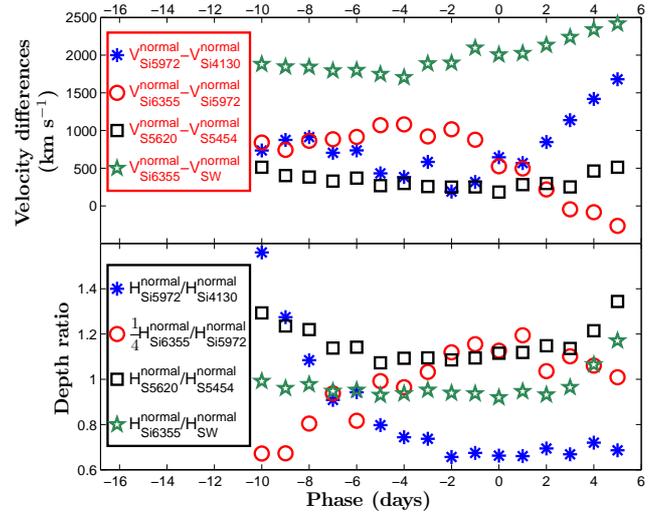} \caption{Upper panel: Velocity differences as a function of phase; Lower panel: Depth ratio as a function of phase. The data were averaged over 2 d intervals. Since some 91T/99aa/91bg-like objects have extreme values of velocity difference or depth ratio, here only normal SNe (i.e., NV or HV SNe) are included. }
\label{Fig_tevo_diff}
\end{figure}

Deeper layers (corresponding to slower layers) of the ejecta tend to have higher temperatures favoring the occupations of highly-excited states. Therefore, lines from highly--excited states are expected to have lower velocities than lines from lowly-excited states. This is confirmed in Figure \ref{Fig_tevo_diff} which shows rapid increases of the velocity-differences after maximum light, and it could be a sign of accelerating separation between ejecta layers with different thermal condition. The only exception is $V_{S5620}-V_{S5454}$, which remains very small, likely due to that these two lines have very close energy levels (see discussion in \S \ref{S2}).

\begin{figure}\centering
\includegraphics[width=\columnwidth]{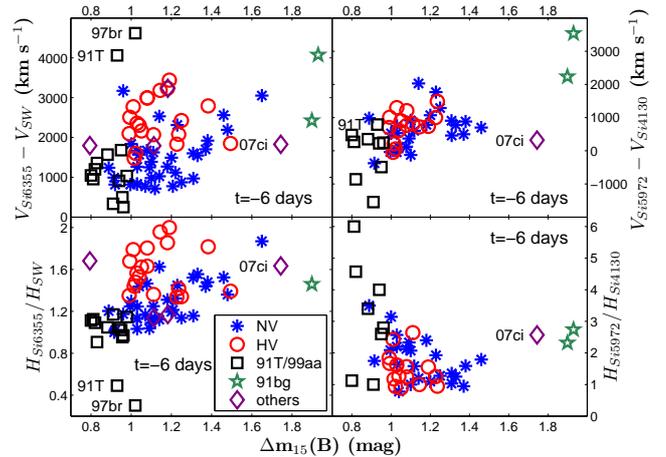} \caption{Velocity-differences and depth-ratios versus $\Delta$m$_{15}$($B$). Measurements within 2 d were used, and if more than one measurement was available, the one closest to the central phases was used. Colors and shapes of data points are the same as in Figure \ref{Fig_V_vs_W}.}
\label{Fig_dm_vs_dVrH}
\end{figure}

Scatter plots in the upper panels of Figure \ref{Fig_dm_vs_dVrH} show details about how $V_{Si6355}-V_{SW}$ and $V_{Si5972}-V_{Si4130}$ are related to $\Delta$m$_{15}$($B$). The sample is restricted to spectra obtained about 6 d prior to maximum brightness, when the velocities decline linearly with time (see Figure \ref{Fig_tevo_V}). The correlations are overall positive, i.e. fast-decliners tend to have greater $V_{Si6355}-V_{SW}$ and $V_{Si5972}-V_{Si4130}$ than slow-decliners \citep[SN 1991T and SN 1997br appear to be two outliers. In fact, they are also considered as peculiar objects in][]{Hachinger06}. The direct reason is that $V_{SW}$ and $V_{Si4130}$ decrease more rapidly with increasing $\Delta$m$_{15}$($B$) than $V_{Si6355}$ and $V_{Si5972}$ (see Figure \ref{Fig_tevo_corr_v}). Physically, this could be explained as that lines with higher excitation and ionization energies are more strongly influenced by the decrease of temperature and luminosity. However, in 91T/99aa-like subclass, $V_{Si5972}-V_{Si4130}$ appears independent of $\Delta$m$_{15}$($B$). This is likely due to that these two lines have similar energy levels, and 91T/99aa-like objects have relatively high temperature and luminosity.

\subsection{Depth ratios\label{S.rw}}

The correlation between the line-strength ratio and $\Delta$m$_{15}$($B$) has been studied before \citep[e.g.,][]{Benetti05,Hachinger06,Blondin12}, with the aim to better calibrate SNe Ia as standard candles, or to trace the element abundances in SN Ia ejecta. Most studies focus on correlations at maximum light, and use pEW-ratios to quantify the strength-ratios. However, as mentioned before, the pEW measurement is easily affected by line-blending, and we therefore use the depth-ratio instead. The other reason we use the depth-ratio is that few of this ratio have been explored. Unless otherwise stated, throughout this work a strength-ratio between two features of the same species is defined as the ratio of the depth of the longer-wavelength feature to the depth of the shorter-wavelength feature.

In Figure \ref{Fig_tevo_diff} we show the time evolution of the depth-ratios. Since some 91T/99aa-like or 91bg-like objects have extreme values of the depth ratios that are not suitable for averaging, this figure includes only normal SNe Ia, i.e. NV or HV groups. It appears that the time evolutions of the depth-ratios share some similarities with the evolutions of the velocity-differences. $H_{Si6355}/H_{SW}$ stays almost unchanged before maximum light ($\approx$ 0.88), suggesting a stable relative-abundance of Si II to S II \citep[see also,][for a comparison between Si II and Ca II]{Zhao15}. $H_{Si5972}/H_{Si4130}$ shows the most significant variation with time, possibly due to a competition between these two lines (see discussion in \S \ref{S.cmp}). 

\begin{figure}\centering
\includegraphics[width=\columnwidth]{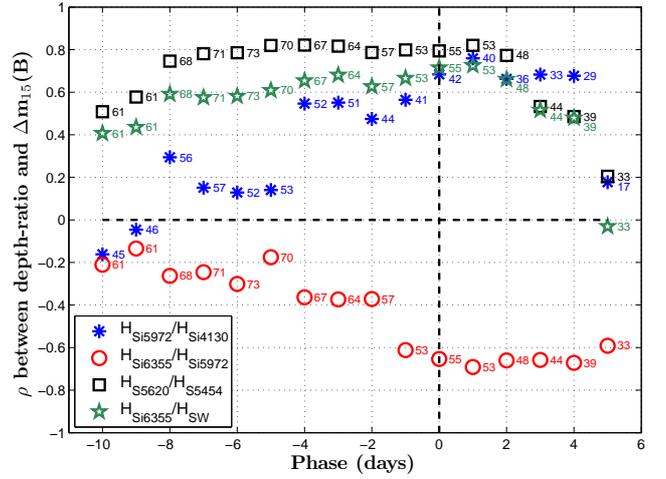}
\caption{Pearson coefficients between depth (``H") ratios and the decline rate $\Delta$m$_{15}$($B$). Measurements within 2 d were used, and if more than one measurement was available, the one closest to the central phases was used. The whole sample, including NV, HV, 91T/99aa-like and 91bg-like SNe are included.}
\label{Fig_tevo_corr_rh}
\end{figure}

\begin{figure}\centering
\includegraphics[width=\columnwidth]{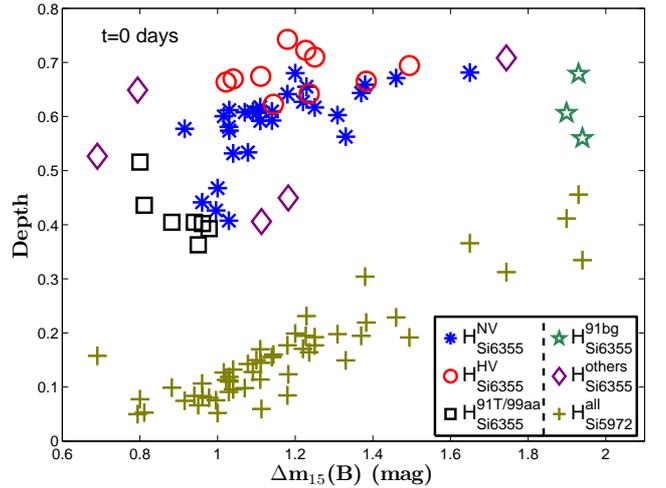} \caption{A comparison of the decline-rate dependences of $H_{Si6355}$ and $H_{Si5972}$. Measurements obtained within 2 d of the maximum light were used, and if more than one measurement was available, the one closest to the specific phase was used.}
\label{Fig_dm_vs_H63555972}
\end{figure}

Fast decliners tend to have lower temperatures \citep[except for some peculiar objects, for example, Iax SNe; see, e.g.,][]{Foley13, Jha17}, favoring the occupation of lowly-excited states. Therefore, a depth ratio of a longer-wavelength feature to a shorter-wavelength feature (eg., $H_{S5620}/H_{S5454}$) is expected to increase with increasing $\Delta$m$_{15}$($B$). This is confirmed in Figure \ref{Fig_tevo_corr_rh}, with $H_{S5620}/H_{S5454}$, $H_{Si5972}/H_{Si4130}$ and $H_{Si6355}/H_{SW}$ showing positive correlations with $\Delta$m$_{15}$($B$) (as indicated by the positive Pearson coefficients). The only exception is $H_{Si6355}/H_{Si5972}$, which is anti-correlated with $\Delta$m$_{15}$($B$), possibly due to high degree of saturation of Si II $\lambda$6355. Line saturation is quite serious for Si II $\lambda$6355 which has rather low energy level. At near- or post-maximum epochs, higher luminosity and lower temperature favor for the formation of absorption lines from lowly excited levels. For example, as shown in Figure \ref{Fig_dm_vs_H63555972}, $H_{Si6355}$ at $B-$band maximum light ceases to grow after reaching an upper limit of about 0.7 at $\Delta$m$_{15}$($B$) $\approx 1.4$ mag. While at $t \approx$ $-6$ d, $pEW_{Si6355}$ still persistently grows with $\Delta$m$_{15}$($B$), as shown in Figure \ref{Fig_dm_vs_W6355SW}.

After maximum light, the correlations between the depth-ratios and $\Delta$m$_{15}$($B$) decrease quickly with time, as shown in Figure \ref{Fig_tevo_corr_rh}. This may be caused by the decreases of temperature and luminosity, and the layers' separations (see discussion in \S \ref{S.dv}). $H_{Si6355}/H_{SW}$ also shows a positive correlation with $\Delta$m$_{15}$($B$), which may be partially due to the fact that dimmer SNe Ia tend to have larger abundance of Si relative to S.

The bottom-left panel of Figure \ref{Fig_dm_vs_dVrH} shows details about how the depth ratio $H_{Si6355}/H_{SW}$ (at $t \approx$ $-6$ d) is related to $\Delta$m$_{15}$($B$). The correlation is overall positive as mentioned above. However, within the 91T/99aa-like subclass, the correlation is reversed, i.e., dimmer objects tend to have smaller $H_{Si6355}/H_{SW}$. This negative correlation results from $H_{Si6355}$'s decreasing more rapidly with $\Delta$m$_{15}$($B$) than $H_{SW}$ in 91T/99aa-like objects, as mentioned in \S \ref{S.wdm}. The physical reason is unclear. But since temperature and luminosity effects are unlikely the reason, we speculate that brighter 91T/99aa-like objects might have more Si-rich materials than dimmer 91T/99aa-like objects (see discussion in \S \ref{S.wdm}).

The bottom-right panel of Figure \ref{Fig_dm_vs_dVrH} shows a complicated correlation between $H_{Si5972}/H_{Si4130}$ at $t \approx$ $-6$ d and $\Delta$m$_{15}$($B$). It appears that there is a positive correlation (i.e., dimmer objects have larger $H_{Si5972}/H_{Si4130}$) for SNe Ia with $\Delta$m$_{15}$($B$) $>1.3$ mag, but a negative correlation at $\Delta$m$_{15}$($B$) $<1.3$ mag. The positive correlation at $\Delta$m$_{15}$($B$) $>1.3$ mag can be explained as that dimmer objects have lower temperatures that would affect Si II $\lambda$4130 more seriously. It is unclear why the correlation is reversed at $\Delta$m$_{15}$($B$) $<1.3$ mag. A possible reason is that brighter objects might have less Si II in the ejecta \citep[as a result of more complete IME burning; see, for example,][]{Benetti05}, and thus an even smaller chance for the absorption of Si II $\lambda$4130 (i.e., a more intense competition between Si II features, see the discussion in \S \ref{S.cmp}).

\subsection{Spectroscopic luminosity indicators\label{S.dmind}}

Correlations between spectroscopic indicators (velocities, pEWs, pEW-ratios etc.) and photometric parameters ($\Delta$m$_{15}$($B$), $B-V$ color etc.) have been extensively studied in attempts to improve the accuracy of SN Ia distance measurements (see references given in \S \ref{S1}). Using a new sample, we reinvestigate the $\Delta$m$_{15}$($B$)-dependences of the line strengths of the four lines studied here. Unlike most previous studies, we use the depth to quantify the line strength, instead of the pEW. This is because, as previously mentioned, the depths generally have stronger correlation with $\Delta$m$_{15}$($B$) than the corresponding pEWs. For example, at $t$ = $-6$, $-3$, 0, and +3 d, the Pearson correlation coefficient $\rho$($pEW_{Si5972}$,$\Delta$m$_{15}$($B$)) = 0.55, 0.71, 0.84, and 0.88, respectively, while $\rho$($H_{Si5972}$,$\Delta$m$_{15}$($B$))= 0.79, 0.85, 0.90, and 0.91, respectively. Also, note that measurement of the depth usually has lower uncertainty than that of the pEW. A pEW-measurement usually involves the whole line profile, and consequently it is more subject to line blending and contamination than the depth.

Figure \ref{Fig_dmInd} shows several spectroscopic indicators as a function of the $\Delta$m$_{15}$($B$). The sample is restricted to NV-SNe whose homogeneous spectroscopic properties can help improve the cosmological distance measurements. At early times, $H_{Si5972}+H_{Si4130}$ shows an even stronger correlation with $\Delta$m$_{15}$($B$) than $H_{Si5972}$ alone, possibly due to an effect of limited element abundance (see \S \ref{S.cmp}). At maximum light, $pEW_{Si5972}$ shows the strongest correlation with $\Delta$m$_{15}$($B$), in line with the results of \citet{Hachinger06} and \citet{Blondin12}. In practice, since it is much easier to obtain a near-maximum spectrum than an early-time spectrum, $H_{Si5972}$ should be more useful than $H_{Si5972}+H_{Si4130}$. $H_{Si4130}$ also has a strong correlation with $\Delta$m$_{15}$($B$), making it another good indicator of the peak luminosity. And, it is particularly useful for high-z SNe Ia \citep{Nordin11}, as line Si II $\lambda$4130 still remains visible in their spectra.

\begin{figure}\centering
\includegraphics[width=\columnwidth]{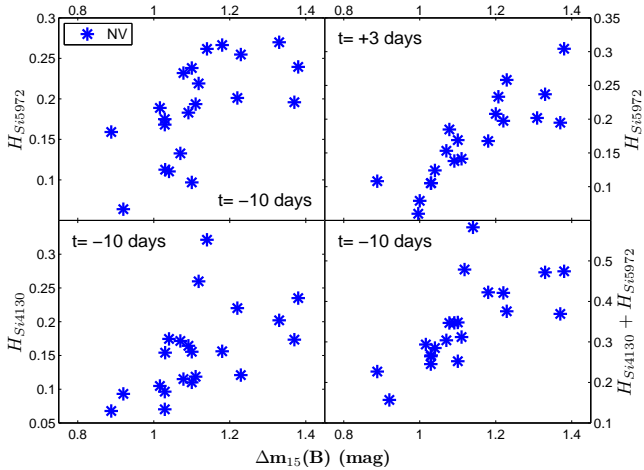} \caption{A comparison of the decline-rate dependences of $H_{Si4130}$ ($-10$ d), $H_{Si5972}$ ($-10$ and +3 d) and $H_{Si4130}+H_{Si5972}$ ($-10$ d). Measurements within 2 d were used, and if more than one measurement was available, the one closest to the central phases was used. The sample is restricted to normal-velocity (NV) objects only.}
\label{Fig_dm_vs_H59724130}
\end{figure}

Scatter plots in Figure \ref{Fig_dm_vs_H59724130} show more clearly how well $H_{Si4130}$, $H_{Si5972}$, and $H_{Si5972}+H_{Si4130}$ correlate with $\Delta$m$_{15}$($B$). It is clear that at $t \approx$ $-10$ d, $H_{Si5972}+H_{Si4130}$ has the tightest relation with $\Delta$m$_{15}$($B$). Figure \ref{Fig_dm_vs_H59724130} also compares $t \approx$ $-10$ d $H_{Si5972}$ correlation with $\Delta$m$_{15}$($B$) to that at $t \approx$ +3 d. It is clear that the correlation is much tighter at $t \approx$ +3 d than at $t \approx$ $-10$ d (see discussions in \S \ref{S.wdm}).

\subsection{Line competition\label{S.cmp}}

Since every element has a limited abundance, there could be a competition of two features from the same element. And, the competition could be more intense when these two features have similar energy levels, as they could be closely related in the radiation transport \citep[e.g.,][]{Sim13,Sim17}. Below we show some spectral behaviors that could be related to line competition.

As noted in \S \ref{S.vt} and \S \ref{S.wt}, Si II $\lambda$5972 behaves quite differently compared to other features. For example, as shown in Figure \ref{Fig_tevo_EW} (here we focus on NV SNe), during the epoch from $-13$ to $-2$ d, Si II $\lambda$5972 weakens with time, while Si II $\lambda$4130 strengthens with time. This is hard to explain with temperature effect or abundance difference, as these two features are both from Si II feature and have very similar energy levels. A possible explanation is that Si II $\lambda$5972 may have serious competition with Si II $\lambda$4130, forcing the later to be located in a much deeper layers. This also explains the fact that at early times Si II $\lambda$5972 has a much higher velocity than Si II $\lambda$4130 (see Figure \ref{Fig_tevo_V}). At maximum light, the competition may be less intense due to relatively high luminosity (and thus more abundant photons), leading to a much smaller velocity difference between Si II $\lambda\lambda$5972 and 4130. Then after maximum light, as the luminosity decreases, the competition may be intensified, causing a rebound of the velocity difference. Similarly, line competition may help explain the great difference between the line strengths of Si II $\lambda\lambda$5972 and 4130. For example, at phases $t \approx$ $-14$ to $-10$ d, $pEW_{Si4130} \approx \frac{1}{4}pEW_{Si5972}$, while near maximum light, $pEW_{Si4130} \approx pEW_{Si5972}$, as shown in Figure \ref{Fig_tevo_EW}. A competition between Si II $\lambda\lambda$5972 and 4130 may also help explain the complicated correlation between strength-ratio $H_{Si5972}/H_{Si4130}$ and $\Delta$m$_{15}$($B$) (see \S \ref{Fig_dm_vs_dVrH}).

A similar case is between the two components of S II W-trough, i.e. S II $\lambda$5454 and S II $\lambda$5620, which originate from almost the same levels ($E_{lower}$: 13.73 eV vs. 13.70 eV, $E_{exc}$: 2.21 eV vs. 2.27 eV). As shown in Figure \ref{Fig_tevo_EW}, before $t \approx$ $-10$ d the redder component has almost twice the strength of the bluer component. While near maximum light, the two components have almost the same strengths. A possible explanation is that there is a competition between the two components. The other possible reason is the saturation effect, as some of the lines from a singly-ionized ion may be saturated for an increasing abundance of the singly-ionized state. 

\section{Conclusions\label{S5}}

The key to improving the cosmological distance measurements with SNe Ia is to correct for their intrinsic diversity as revealed in the spectral diversity. The aim of this paper is to better understand the physics of the SN Ia features, which would provide a basis for quantitative correction of the intrinsic diversity. We examined four important features of Si II and S II for 554 spectra of 76 SNe Ia, and investigated the time evolutions of their velocities and line strengths, their velocity-differences and depth-ratios, their correlations with $\Delta$m$_{15}$($B$), and the diversity among the subclasses. Below is a summary of our major findings:

1. \emph{Velocity:} Line velocities generally decrease with time as expected, but an exception is the velocity of Si II $\lambda$5972, which somehow increases after $-2$ d and even surpasses the velocity of Si II $\lambda$6355 after +2 d. Faster-decliners tend to have faster declining velocities, possibly resulting from faster decrease of temperature.

2. \emph{Line strength:} Before $-7$ d, line strengths of the two longer-wavelength features, i.e. Si II $\lambda\lambda$5972 and 6355 decrease with time, while line strengths of the two shorter-wavelength features, i.e. Si II $\lambda$4130 and S II trough, increase with time. After $-7$ d, the line strengths of all features increase with time, and reach a flat maxima or plateau near maximum light, except the strength of Si II $\lambda$5972 which reaches a minima near maximum light. A possible reason for the mysterious behavior of Si II $\lambda$5972 might be its competition with other Si II features.

3. \emph{Velocity-pEW correlation}: The overall correlation between velocity and pEW is positive. But for different subclasses at different phases, the correlation can be very different. For example, for HV and 91T/99aa-like objects at $-6$ d, $V_{Si6355}$ appears to be positively correlated with $pEW_{Si6355}$. While for NV objects at +3 d, $V_{Si6355}$ appears to be anti-correlated with $pEW_{Si6355}$.

4. \emph{$\Delta$m$_{15}$($B$)-dependence of the velocity}: There is an overall trend of increasing velocity for the brighter SNe. But this trend is only significant near +3 d, and is almost nonexistent within NV subclass.

5. \emph{$\Delta$m$_{15}$($B$)-dependence of the line strength}: Among the NV SNe, there is a trend of stronger Si II and S II features for faster decliners, possibly a result of lower degree of ionization and more remained IMEs in the ejecta of dimmer ones. Among the 91T/99aa-like SNe, on the contrary, slower decliners appear to have stronger Si II and S II absorptions.

6. \emph{Velocity-difference:} There is an overall trend of larger $V_{Si6355}-V_{SW}$ for fainter objects at early times. But within 91T/99aa-like SNe the trend is reversed, possibly due to an effect of ionization. The absolute values of the velocity-differences between different features increase with time after maximum light, suggesting an accelerating separation of the ejecta layers.

7. \emph{Depth-ratio:} Depth-ratios of a longer wavelength feature to a shorter wavelength feature show positive correlations with $\Delta$m$_{15}$($B$), though the correlations weaken quickly with time after maximum light. The only exception is the $H_{Si6355}/H_{Si5972}$, which is anti-correlated with $\Delta$m$_{15}$($B$), possibly due to the saturation of Si II $\lambda$6355. Time-evolutions of the depth-ratios are slower than the evolutions of velocity-differences, especially for $H_{Si6355}/H_{SW}$, suggesting relatively stable element abundances in the ejecta.

8. \emph{Luminosity indicator}: Less affected by line-blendings, the absorption depths may be better indicators of the luminosity than the corresponding pEWs. Near maximum light, $H_{Si5972}$ shows the strongest correlation with $\Delta$m$_{15}$($B$), while at early times $H_{Si4130}+H_{Si5972}$ shows the strongest correlation with $\Delta$m$_{15}$($B$).

The results of this study are very suggestive, but require further investigation. This would require an even larger spectral sample that covers extensive observations of all subtypes. The results of this paper may also help constrain the explosion models (e.g., help determine the abundances of burning products at different phases through the line strengths), or be used to improve the accuracy of SNe Ia as distance indicators. Further theoretical analysis or modeling would be very helpful to better understand the results we obtain here.

\section*{Acknowledgements}
We thank the anonymous referee for his/her insightful suggestions which help improve the paper a lot. The work of X. Wang is supported by National Science Foundation of China (12033003, 11633002, and 11761141001), and the National Program on Key Research and Development Project (grant No. 2016YFA0400803). K.M. acknowledges support provided by Japan Society for the Promotion of Science (JSPS) through KAKENHI grant (17H02864, 18H04585, 18H05223, 20H00174, and 20H04737). This research has made use of the CfA Supernova Archive, which is funded in part by the US National Science Foundation through grant AST 0907903. This research has also made use of Berkeley/Lick Supernova Archives, which is funded in part by the US National Science Foundation. This research has also made use of the CSP Supernova Archive, which is supported by the World Premier International Research Center Initiative (WPI).

\section*{Data availability}
The authors confirm that the data supporting the findings of this study are available within the article and its supplementary materials.


\bsp	
\label{lastpage}


\end{document}